\documentclass[12pt]{article}
\usepackage{amsmath,amssymb,mathtools,amsfonts,epsfig,graphicx,euscript}%,mathrsfs}
 \usepackage{color}
\usepackage{dsfont}
\usepackage{etoolbox}
\usepackage{xcolor}
\usepackage{enumitem}
\usepackage{multirow}
\usepackage{float}
\usepackage{slashed}
\usepackage{appendix}
\usepackage{subfig}
\usepackage[compat=1.1.0]{tikz-feynman}
\begin{document}
\begin{titlepage}
\thispagestyle{empty}

\vspace{4cm}
\begin{center}
\font\titlerm=cmr10 scaled\magstep4
\font\titlei=cmmi10
scaled\magstep4 \font\titleis=cmmi7 scaled\magstep4 {\Large{\textbf{Matter and dark matter asymmetry from a composite Higgs model}
\\}}
\setcounter{footnote}{0}
\vspace{1.5cm} 
\noindent{{M. Ahmadvand \footnote{e-mail:ahmadvand@ipm.ir} 
		}}\\
\vspace{0.2cm}

{\it School of Physics, Institute for Research in Fundamental Sciences (IPM), P. O. Box 19395-5531, Tehran, Iran\\}

\vspace*{.4cm}
\end{center}

\vskip 2em
\setcounter{footnote}{0}
%----------------------------------------------------------------------------
\begin{abstract}
We propose a low scale leptogenesis scenario in the framework of composite Higgs models supplemented with singlet heavy neutrinos. One of the neutrinos can also be considered as a dark matter candidate whose stability is guaranteed by a discrete $\mathbb{Z}_2 $ symmetry of the model. In the spectrum of the strongly coupled system, bound states heavier than the pseudo Nambu-Goldstone Higgs boson can exist. Due to the decay of these states to heavy right-handed neutrinos, an asymmetry in the visible and dark sector is simultaneously generated. The resulting asymmetry is transferred to the standard model leptons which interact with visible right-handed neutrinos. We show that the sphaleron-induced baryon asymmetry can be provided at the TeV scale for resonant bound states. Depending on the coupling strength of dark neutrino interaction, a viable range of the dark matter mass is allowed in the model. Furthermore, taking into account the effective interactions of dark matter, we discuss low-energy processes and experiments.

\end{abstract}
\end{titlepage}
%%%%%%%%%%%%%%%%%%%%%%%%%%%%

\baselineskip=1.3\baselineskip	

\section{Introduction}
The Standard Model (SM) as a gauge theory for elementary particle interactions has been extremely successful in explaining many phenomena. However, some issues including the matter-antimatter asymmetry of the Universe, the Dark Matter (DM) and hierarchy problem have remained unresolved \cite{Riotto:1999yt,Bertone:2004pz,Giudice:2008bi}. These shortcomings imply that the SM is incomplete and needs to be extended. 

Astrophysical evidence implies the Universe is asymmetric with an overabundance of baryons relative to antibaryons \cite{Cohen:1997ac}. Based on cosmological abundances of light nuclei and also CMB observations \cite{Zyla:2020zbs}, one can characterize the baryon asymmetry by this ratio $ n_{\mathrm{B}}/s\sim 0.88\times 10^{-10} $ where $ n_{\mathrm{B}}$ is the net baryon number density and $ s$ is the entropy density of the Universe. Starting with an initial symmetric state of matter and antimatter, this ratio can be obtained in the scenario which provides three conditions \cite{Sakharov:1967dj}: baryon number violation, C and CP violation, and departure from thermal equilibrium. To achieve the baryon asymmetry, various scenarios beyond the SM, such as GUT baryogenesis \cite{Weinberg:1979bt, Riotto:1998bt}, Electroweak (EW) baryogenesis \cite{Trodden:1998ym, Ahmadvand:2013sna}, and leptogenesis which was first proposed in the seesaw mechanism context \cite{Fukugita:1986hr, Davidson:2008bu}, have been suggested so far.

On the other hand, recent observations show that around 26\% of the energy of the Universe lies in DM  \cite{Zyla:2020zbs}. However, the nature of DM and its properties such as its mass and spin have not been revealed. To obtain the relic density of DM, numerous models with DM candidates have been proposed among which we can enumerate weakly interacting massive particles (WIMPs), sterile neutrinos, axions, and primordial black holes \cite{Lin:2019uvt}. As for this problem, another notable observation is that the energy density of DM is so close to that of baryonic matter, $ \Omega_{\mathrm{DM}}\sim5\Omega_{\mathrm{B}} $. This relation may hint the dark and visible matter have a common asymmetric origin \cite{Kaplan:1991ah}.

Another problem which cannot be addressed by the SM is the hierarchy problem, concerning the large corrections of UV physics to the mass of the Higgs as an elementary particle. One of the attractive solutions is Composite Higgs Models (CHMs) in which Higgs is no longer an elementary particle but a composite state of a new strongly coupled sector \cite{Kaplan:1983sm, Panico:2015jxa}. Indeed, the Higgs is considered as a pseudo Nambu-Goldstone boson, resulted from a spontaneously broken global symmetry, so that there is a mass gap between the Higgs state and other heavier strongly interacting bound states.

In this paper, to address the aforementioned problems, we propose a leptogenesis scenario through the possibilities of a CHM. We consider a minimal CHM supplemented with at least two singlet heavy neutrinos, where one of them can be regarded as a DM candidate. Also, in the spectrum of the strongly coupled system, large number of bound states are expected, analogous to QCD hadrons. We consider bound states which decay to Right-Handed (RH) heavy neutrinos. Due to their complex couplings, leading to a CP violation in the decay processes, the asymmetry can be generated at the compositeness scale, about TeV scale \cite{Panico:2015jxa}. (Thus, we assume bound states are constituted of heavy so-called techniquarks.) Moreover, because of the interaction of visible RH neutrinos with SM leptons, the asymmetry is transferred to this sector. Then, the baryon asymmetry is generated through active EW sphalerons\footnote{For the sphaleron energy calculation in the context of CHMs, see \cite{Spannowsky:2016ile}.}. Depending on the model parameters, for resonant bound states, we obtain the observed quantities and also find a valid range of DM masses. Eventually, we discuss relevant effective interactions of dark matter with SM particles at low-energy scales.

Therefore, as the Higgs state interactions can be responsible for the mass generation of SM particles and the spontaneous gauge symmetry breaking, the asymmetric matter and DM may be originated from the decay of some other bound states.

In section 2 we introduce the model and obtain the parameters required for the asymmetric model. In section 3 we represent numerical examples and calculate relevant parameters. We conclude in section 4.
	
\section{Model}
Depending on the global symmetry coset, CHMs can be classified. In the minimal version with the coset $ \mathrm{SO(5)/SO(4)}$, one Nambu-Goldstone (NG) Higgs doublet is generated due to the symmetry breaking and also the SM gauge symmetry is a subgroup of the unbroken $ \mathrm{SO(4)}$ symmetry group. In fact, the Higgs is a pseudo NG boson and the loop-induced composite Higgs potential is generated from the explicit breaking of the Goldstone symmetry, through elementary-composite interactions. Because of the pseudo NG boson nature of the Higgs, the generic form of the potential has a trigonometric structure as \cite{Panico:2015jxa}
\begin{equation}
V(H)=-\alpha_0 f^2\sin ^2\frac{H}{f}+\beta_0 f^2\sin ^4\frac{H}{f} 
\end{equation}
where $ H$ is the Higgs doublet, $ f$ denotes the compositeness scale and $ (\alpha_0, \beta_0)$ parameters encode fermion and gauge sector contributions. To obtain realistic EW symmetry breaking scale and Higgs mass, $ \alpha_0=2\beta_0\xi $ and $ m_h^2=8\xi(1-\xi)\beta_0 $ where $ \xi=\sin^2(\langle H\rangle/f)=\upsilon^2/f^2 $ and $\upsilon=246\,\mathrm{GeV} $. In addition, to satisfy EW precision tests and Higgs coupling measurements, $ \xi\lesssim 0.1 $ \cite{Grojean:2013qca}. We take the compositeness scale $ f\sim 1\,\mathrm{TeV}$, which is a case without fine tuning.

In this context, in order to introduce the interaction of the Higgs with SM fermions and also to generate their masses, the idea of partial compositeness \cite{Kaplan:1991dc} is expressed as $ Q_{\mathrm{SM}}\mathcal{O}_f $ where $ Q_{\mathrm{SM}} $ stands for the SM fermions which couple to the composite sector through $\mathcal{O}_f $ fermionic operator. For example, the interaction of SM quarks can be written as $\lambda_u \bar{u}_R\mathcal{O}_5^R+\cdots $ (plus analogous terms for $ q_L$ and other flavors). However, the SM fields should be embedded in $ \mathrm{SO(5)}$ representation, that is to say $\lambda_u \overline{U}_R^{\mathbf{5}}\mathcal{O}_{\mathbf{5}}^R+\cdots $, where $ U_R^{\mathbf{5}}=(0, 0, 0, 0, u_R)^T $ is the fiveplet of $ \mathrm{SO(5)}$ and the $ T$ symbol denotes the transpose. We can employ the Callan-Coleman-Wess-Zumino (CCWZ) construction \cite{Coleman:1969sm,Callan:1969sn} and split $ \mathrm{SO(5)} $ multiplets into those of $ \mathrm{SO(4)} $,
\begin{equation}\label{dr1}
(U^{\mathbf{4}}_R, U^{\mathbf{1}}_R)^{T}=U[\Pi]^{-1}\cdot U_R^{\mathbf{5}} 
\end{equation}
where the Goldstone matrix inverse is given by \cite{Panico:2015jxa}
\begin{equation}\label{gm1}
	U[\Pi]^{-1}=\begin{pmatrix}
		\mathds{1}-\Big(1-\cos\frac{\Pi}{f}\Big)\frac{\mathbf{\Pi}\cdot\mathbf{\Pi}^T}{\Pi^2} & -\frac{\mathbf{\Pi}}{\Pi}\sin\frac{\Pi}{f}  \\
		\frac{\mathbf{\Pi}^T}{\Pi}\sin\frac{\Pi}{f}  & \cos\frac{\Pi}{f}
	\end{pmatrix} 
\end{equation}
$ \Pi=\sqrt{\mathbf{\Pi}^T\cdot\mathbf{\Pi}} $ and the Higgs can be defined by components of the Goldstone vector, $ \mathbf{\Pi}$. Thus, from Eqs.\ (\ref{dr1}, \ref{gm1}),
\begin{equation}
U^{\mathbf{4}}_R=-U_R\frac{\mathbf{\Pi}}{\Pi}\sin\frac{\Pi}{f},~~~~~~~ U^{\mathbf{1}}_R=U_R\cos\frac{\Pi}{f}.
\end{equation}
On the other hand, the multiplets should contain the representations of SM fermions. Indeed, when the EW symmetry group is included in $ \mathrm{SO(5)} $, the hypercharge of SM fermions is not reproduced. Therefore, an extension of the global symmetry is required, $\mathrm{SO(5)}\times \mathrm{U(1)}_X\rightarrow \mathrm{SO(4)}\times \mathrm{U(1)}_X $ \cite{Panico:2015jxa}, so that the hypercharge of fermions is now defined as $ Y=T_R^3+X$. The hypercharge of $ u_R$ singlet is 2/3, hence by choosing $X$ charge 2/3, we have $ \mathbf{5}_{2/3}\rightarrow\mathbf{4}_{2/3}\oplus\mathbf{1}_{2/3}\rightarrow\mathbf{2}_{7/6}\otimes\mathbf{2}_{1/6}\otimes\mathbf{1}_{2/3} $, where $\mathbf{2}_{1/6} $ is identified with the LH quark doublet $ q_L$ and the same procedure holds for it. The doublet is embedded in $ Q_{u_L}^{\mathbf{5}}=(\boldsymbol{q}_{L_{-}}, 0)^T $. The Fourplet can be generally expressed in terms of components of $ \mathrm{SU(2)}_L$ doublets, $ \Psi_{\mp} $, with hypercharge $\mp 1/2$ as
\begin{equation}
\frac{1}{\sqrt{2}}\left(-i \Psi_{+}^{u}-i \Psi_{-}^{d}, \Psi_{+}^{u}-\Psi_{-}^{d}, i \Psi_{+}^{d}-i \Psi_{-}^{u}, \Psi_{+}^{d}+\Psi_{-}^{u}\right)^{T}.
\end{equation}
Because $q_L$ has hypercharge 1/6, for $ X=2/3$, the projected doublet would have  $ T_R^3=-1/2$, i.e., $ q_L=\Psi_- (\Psi_+=\mathbf{0})$. Thus,
\begin{equation}
\boldsymbol{q}_{L_{-}}=\frac{1}{\sqrt{2}}\left(-i d_L, -d_L, -i u_L, u_L\right)^{T}.
\end{equation}
Now, we can dress $ Q_{u_L}^{\mathbf{5}}$ similar to Eq.\ (\ref{dr1}) so that 
\begin{equation}
Q_{u_{L}}^{\mathbf{4}}=\left(\mathds{1}-\Big(1-\cos \frac{\Pi}{f}\Big) \frac{\mathbf{\Pi}\cdot\mathbf{\Pi}^T}{\Pi^{2}}\right) \cdot \boldsymbol{q}_{L_{-}}, \quad Q_{u_{L}}^{\mathbf{1}}=\left(\frac{\mathbf{\Pi}^T}{\Pi} \sin \frac{\Pi}{f}\right) \cdot \boldsymbol{q}_{L_{-}}.
\end{equation}
The invariant can be attained from $ \overline{Q}^{\mathbf{1}}_{u_L} U_R^{\mathbf{1}} $, which is equivalent to $ \overline{Q}^{\mathbf{4}}_{u_L} U_R^{\mathbf{4}} $. As a result, the generalized Yukawa interaction can be obtained from
\begin{equation}\label{nlh}
\overline{Q}^{\mathbf{1}}_{u_L} U_R^{\mathbf{1}}= \frac{1}{2\Pi}\sin\frac{2\Pi}{f}\overline{q}_{L}H^c u_{R}= \frac{1}{2\sqrt{2}|H|}\sin\frac{2\sqrt{2}|H|}{f}\overline{q}_{L}H^c u_{R}.
\end{equation}
By setting the Higgs to its vacuum, $V$, and Taylor-expanding $ \sin \frac{2(V+h)}{f}$ around $ h=0$, where $h$ is the physical Higgs field,
\begin{equation}
\begin{aligned}
	\sin \frac{2(V+h)}{f} & \simeq 2 \sin \frac{V}{f} \cos \frac{V}{f}+\frac{2 h}{f}\Big(1-2 \sin ^{2} \frac{V}{f}\Big)-\frac{4}{f^{2}} \sin \frac{V}{f} \cos \frac{V}{f}+\cdots \\
	& \simeq 2 \sqrt{\xi(1-\xi)}+2 \frac{h}{\upsilon}(1-2 \xi) \sqrt{\xi}-4 \frac{h^{2}}{\upsilon ^{2}} \sqrt{\xi(1-\xi)} \xi+\cdots,
\end{aligned}
\end{equation}
one can find for example the modified Yukawa coupling as 
\begin{equation}
\frac{g^\mathrm{comp}_{huu}}{g^\mathrm{SM}_{huu}}=\frac{1-2 \xi}{\sqrt{1-\xi}}
\end{equation}
where such deviations are allowed by the LHC data \cite{Aad:2015gba,Khachatryan:2014jba} for $ \xi\lesssim 0.1 $.

In the paper, we describe the scenario in the confined phase. For the sake of simplicity, we take into account two given bound states as gauge singlet scalar fields. Moreover, we consider two gauge singlet heavy neutrinos with their chiral components. The interaction between these two sectors, actually between the bound states and RH neutrinos, is a lepton-number violating process. Therefore, the Lagrangian is given as follows\footnote{A Majorana mass term can be generated in the model but we assume it is smaller than the Dirac one  \cite{Agashe:2018oyk}.} 
\begin{equation}\label{la}
\mathcal{L}\supset \frac{1}{2}\partial_{\mu}\varphi_i\partial^{\mu}\varphi_i-\frac{1}{2} M_i^2\varphi_i^2+\overline{N}_{R,\alpha}i\slashed{\partial}N_{R,\alpha}-m_{N_{\alpha}}\overline{N}_{L,\alpha}N_{R,\alpha}+Y_{\alpha\alpha}^{i}\,\varphi _i\overline{N}_{R, \alpha}N_{R,\alpha}^c+\mathrm{h.c.}.
\end{equation}
We consider the lepton-number of RH neutrinos as $ L(N_R)=1$ with the opposite sign for their antiparticles and hence the interaction gives rise to a lepton-number violation. $N_{R,2} $ is a $ \mathbb{Z}_2$-odd DM candidate, while other fields are even under this symmetry.

Bound states and heavy neutrinos are taken as singlets of the unbroken global symmetry. Fields of the model can be embedded in multiplets of the unbroken $ \mathrm{SO(4)}$ group. In order for the Lagrangian to be invariant under the symmetry group, we employ the CCWZ construction. For instance for (RH) heavy neutrinos, this would be (flavor indices are not shown for brevity)
\begin{equation}\label{dr}
(N^{\mathbf{4}}_R, N^{\mathbf{1}}_R)^{T}=U[\Pi]^{-1}\cdot N_R^{\mathbf{5}} 
\end{equation}
where the embedding of $ N_R$ in the fiveplet of $ \mathrm{SO(5)}$ is $ N_R^{\mathbf{5}}=(0, 0, 0, 0, N_R)^T $, and $N^{\mathbf{4}}_R $ and $ N^{\mathbf{1}}_R$ belong to $ \mathbf{4}$ and $ \mathbf{1}$ representation of $ \mathrm{SO(4)}$. Thus,
\begin{equation}
N^{\mathbf{4}}_R=-N_R\frac{\mathbf{\Pi}}{\Pi}\sin\frac{\Pi}{f},~~~~~~~ N^{\mathbf{1}}_R=N_R\cos\frac{\Pi}{f}.
\end{equation}
The above Yukawa interaction term, Eq.\ (\ref{la}), can be also generated by the following invariant\footnote{This term may be also originated from a four-fermion interaction $\overline{\psi}_R\psi_L\overline{N}_{R, \alpha}N_{R,\alpha}^c $, where $\psi_{L, R} $ techni-quark fields are assumed to be singlets of $ \mathrm{SO(5)}$ and constituents of $ \varphi_{i}$ bound states.}
\begin{equation}
\mathcal{L}_{\mathrm{int}}=Y_i \overline{N}_R^{\mathbf{5}}\mathcal{O}_{\mathbf{5}}+h.c.=Y_i(\overline{N}_R^{\mathbf{4}}\mathcal{O}_{\mathbf{4}}+\overline{N}_R^{\mathbf{1}}\mathcal{O}_{\mathbf{1}})+h.c.
\end{equation}
where
\begin{equation}
\mathcal{O}_{\mathbf{4}}=-\varphi_i N_R^c\frac{\mathbf{\Pi}}{\Pi}\sin\frac{\Pi}{f},~~~~~~~ \mathcal{O}_{\mathbf{1}}=\varphi_i N_R^c\cos\frac{\Pi}{f}
\end{equation}
and $ \varphi_i$ bound states as singlets of $ \mathrm{SO(4)}$ can be individually introduced instead of grouping them in a $ \mathbf{5}$ multiplet. In addition, the interaction of $N_{R,1} $ with SM leptons is necessary. This can be also fulfilled via partial compositeness paradigm from singlets, $ \overline{E}^{\mathbf{1}}_L N_R^{\mathbf{1}} $, and the SM LH leptons can be embedded in $E^{\mathbf{1}}_L $. Since $ \Psi_{L}=(v_{L}, e_{L})^T$ as the SM LH lepton doublet has $ Y=-1/2$, for $ T_R^3=-1/2$
\begin{equation}
E^{\mathbf{1}}_L=\Big(\frac{\mathbf{\Pi}^T}{\Pi}\sin\frac{\Pi}{f}\Big)\cdot\Psi_{L_-},~~~~~~~ \Psi_{L_-}=\frac{1}{\sqrt{2}}(-ie_L, -e_L, -iv_L, v_L)^T
\end{equation}
where for the $2\times 2$ matrix in this case $ \Psi\equiv(\Psi_-=\Psi_L, \Psi_+=\mathbf{0})$. Therefore, the required interaction can be produced by
\begin{equation}\label{nlh}
\overline{E}^{\mathbf{1}}_L N_R^{\mathbf{1}}= \frac{1}{2\Pi}\sin\frac{2\Pi}{f}\overline{\Psi}_{L,l}H^c N_{R,1}= \frac{1}{2\sqrt{2}|H|}\sin\frac{2\sqrt{2}|H|}{f}\overline{\Psi}_{L,l}H^c N_{R,1}.
\end{equation}

\subsection{L and CP violation}
According to the interaction term, we can obtain the decay rate at the tree level as 
\begin{equation}
\Gamma_i\equiv \sum_{\alpha} \Gamma _{\alpha\alpha}^i (\varphi_i\rightarrow N_{R, \alpha}N_{R, \alpha})=\frac{(Y^{\dag}Y)_{ii}M_{i}}{16\pi} 
\end{equation}
where the process generates a lepton asymmetry with $ \Delta L=2$. However, in order to gain a net asymmetry, there should be enough CP violation in the model. In our scenario, this can be provided through the introduced complex couplings.

To obtain the CP asymmetry in $\varphi_i$ decays, we calculate the interference of tree  and one-loop level amplitudes through the following quantity
\begin{equation}
\varepsilon ^i_{\alpha\alpha}=\frac{\Gamma ^i_{\alpha\alpha}-(\Gamma ^i_{\alpha\alpha})^c}{\Gamma _{i}+\Gamma _{i}^c} 
\end{equation}
where $c$ denotes the CP conjugate of the decay process, $\varphi_i\rightarrow \overline{N}_{R,\alpha}\overline{N}_{R, \beta} $. At the one-loop level, self-energy and vertex diagrams contribute to the CP asymmetry. Since we are dealing with unstable states which cannot be described as asymptotic states by the conventional perturbation field theory, we use the resummation approach \cite{Pilaftsis:1997jf, Pilaftsis:2003gt}. By focusing on the self-energy transition, $ \varphi_{i}\rightarrow \varphi_{j}$, the transition amplitude is given by
\begin{equation}
\mathcal{T}_{\varphi_{i}}=Y^i_{\alpha\alpha}\bar{u}_R v_R^c-iY^j_{\alpha\alpha}\frac{\bar{u}_R v_R^c \Pi_{ji}(M_{i}^2)}{M_{i}^2-M_{j}^2+i\Pi_{jj}(M_{i}^2)},~~~~~i, j=1,2,~~~i\neq j 
\end{equation}
where flavor indices of spinors are not displayed for brevity. The absorptive part of the one-loop transition is expressed as
\begin{equation}
\Pi_{ij}(M_{i}^2)=\frac{(Y Y^*)_{ij}}{16\pi}M_{i}^2=A_{ij}M_{i}^2.
\end{equation}
Therefore, the transition amplitude and its CP conjugate will be
\begin{equation}
\mathcal{T}_{\varphi_{i}}=\bar{u}_R v_R^c\Bigg[Y^i_{\alpha\alpha}-iY^j_{\alpha\alpha} \frac{ A_{ji}M_{i}^2\Big(\Delta M_{ij}^2-iM_{i}^2A_{jj}\Big)}{M_{i}^2\Big(1+iA_{jj}\Big)-M_{j}^2}\Bigg],
\end{equation}
\begin{equation}
\mathcal{\overline{T}}_{\varphi_{i}}=\bar{u}_R^c v_R\Bigg[(Y^i_{\alpha\alpha})^*-i(Y^j_{\alpha\alpha})^*\frac{ A_{ji}^*M_{i}^2\Big(\Delta M_{ij}^2-iM_{i}^2A_{jj}\Big)}{M_{i}^2\Big(1+iA_{jj}\Big)-M_{j}^2}\Bigg]
\end{equation}	
where $\Delta M_{ij}^2\equiv M_{i}^2-M_{j}^2 $. Thus, the CP asymmetry quantity is obtained as
\begin{equation}\label{cp}
\varepsilon ^i_{\alpha\alpha}=\frac{\mathrm{Im}[(Y^{\dag}_{\alpha\alpha} Y_{\alpha\alpha})^2_{ij}]}{8\pi (Y^{\dag} Y)_{ii}}\frac{M_{i}M_{j}(\Delta M_{ij}^2)}{(\Delta M_{ij}^2)^2+M_{i}^2M_{j}^2A^2_{jj}}=\frac{\mathrm{Im}[(Y^{\dag}_{\alpha\alpha} Y_{\alpha\alpha})^2_{ij}]}{(Y^{\dag} Y)_{ii}(Y^{\dag} Y)_{jj}}\frac{M_{i}\Gamma_{j}(\Delta M_{ij}^2)}{(\Delta M_{ij}^2)^2+M_{i}^2\Gamma^2_{j}}.  
\end{equation}	
Satisfying the condition $ M_{i}-M_{j}=\Gamma_{j}/2 $, the CP asymmetry can be $ |\varepsilon ^i_{\alpha\alpha}|\leq 1/2 $. In this case, we can neglect the vertex contribution. Additionally, resonant states may be implied to CP invariant bound states which are a mixture of $ \varphi_1$ and $ \varphi_2$, analogous to the neutral koan system in QCD \cite{Griffiths:2008zz}.

\subsection{Boltzmann equations}

To obtain the number density of particles involving in the out of equilibrium processes, one can solve Boltzmann equations. We consider only decays and inverse decays of $\varphi_i $ fields which dominantly contribute to the creation and washout of the asymmetry. Therefore, the Boltzmann equations up to $ \mathcal{O}(Y^2)$ are expressed as follows\footnote{At $\mathcal{O}(Y^4) $, $ \Delta L=2$ t-channel scattering, $ \varphi\varphi\leftrightarrow N_R N_R $ is induced.}
\begin{equation}\label{b1}
\frac{dy_{\varphi_{i}}}{dz}=\frac{-2z\,\gamma^{\mathrm{eq}}_{\varphi_{i}}}{H_1 s\,y^{\mathrm{eq}}_{\varphi_{i}}}\Big(y_{\varphi_{i}}-y_{\varphi_{i}}^{\mathrm{eq}}\Big)
\end{equation}
\begin{equation}\label{b2}
\frac{d\mathcal{L}_{\alpha}}{dz}=\frac{2z}{H_1s}\Big[\frac{\varepsilon^i _{\alpha\alpha}}{y^{\mathrm{eq}}_{\varphi_{i}}}\Big(y_{\varphi_{i}}-y_{\varphi_{i}}^{\mathrm{eq}}\Big)\gamma^{\mathrm{eq}}_{\varphi_{i}}-\frac{\mathcal{L}_{\alpha}}{y^{\mathrm{eq}}_{L_{\alpha}}}\gamma^{\mathrm{eq}}_{L_{\alpha}}\Big]
\end{equation}
where $z=M_i/T $, $ H_1\equiv H(z=1)=1.66\sqrt{g_*}M^2_i/M_{\mathrm{pl}}$ is the Hubble parameter, $s=2\pi^2g_*T^3/45 $ is the entropy density, $ M_\mathrm{pl}=1.22\times 10^{19}\,\mathrm{GeV}$ is the Planck mass, and $g_*$ denotes the number of relativistic degrees of freedom at temperature $ T$. We also defined each density as $  y_X\equiv n_X/s$ and $ \mathcal{L}_{\alpha}\equiv y_{N_{\alpha}}-y_{\bar{N}_{\alpha}}$. In Eqs.\ (\ref{b1}, \ref{b2}), the thermally averaged rates are given as \cite{Davidson:2008bu, Giudice:2003jh}
\begin{equation}
\frac{\gamma^{\mathrm{eq}}_{\varphi_{i}}}{n^{\mathrm{eq}}_{\varphi_{i}}}=\frac{K_1(z)}{K_2(z)}\Gamma_i ,~~~~~~~~\frac{\gamma ^{\mathrm{eq}}_{L_{\alpha}}}{n^{\mathrm{eq}}_{L_{\alpha}}}=\frac{z^2K_1(z)}{4\pi^2}\Gamma _{\alpha\alpha}^i.
\end{equation}
$K_1(z), K_2(z)$ are modified Bessel functions of the second kind. In the above equations for $\alpha=1$, beside $\Gamma ^i_{11} $, the decay rate resulted from $ N_{R,1}$ decay to the Higgs and SM leptons can be added. However, we can neglect its influence in comparison with $ \Gamma _{11}^i$. 

For large $ z$ in the strong regime \cite{Davidson:2008bu, Buchmuller:2004nz} where the final result is not sensitive to the initial abundance of $ \varphi_{i}$, we can solve the equations analytically and find the asymmetry as
\begin{equation}
\frac{d(y_{\varphi_{i}}-y_{\varphi_{i}}^{\mathrm{eq}})}{dz}\simeq 0,~~~~~~~y_{\varphi_{i}}-y_{\varphi_{i}}^{\mathrm{eq}}\simeq \frac{zK_2(z)H_1}{4g_*\Gamma_i}
\end{equation}
\begin{equation}\label{lep}
\mathcal{L}_{\alpha}\simeq \varepsilon ^i_{\alpha\alpha}\int_{z_i}^{\infty} dz \frac{z^2K_1(z)}{2g_*}\exp\Big(-\int_z^{\infty}dz'\,z'^3 K_1(z')\frac{\Gamma ^i_{\alpha\alpha}}{2H_1}\Big)\equiv\varepsilon ^i_{\alpha\alpha}\eta _{i\alpha}.
\end{equation}
Eventually, the resulting asymmetry of RH neutrinos is transported to the SM leptons through $ N_{R, 1}$ decays, Eq.\ (\ref{nlh}), such that more RH heavy neutrino decays lead to an excess of LH light neutrino and then due to sphalerons which are active at the compositeness scale, the asymmetry is converted to the baryon asymmetry, $ \mathcal{B}\approx -1/3 \mathcal{L}_1$ \cite{Davidson:2008bu}.

\section{Numerical examples}
Having constructed a common origin of matter and DM asymmetry based on $ \varphi_{i}$ decays, we can calculate the observed parameters. In fact, based on the baryon asymmetry and $ \Omega_{\mathrm{DM}}\sim5\Omega_{\mathrm{B}} $ relations, parameters of the model can be constrained by the following numerical analysis.

Firstly, masses of $ \varphi_{i}$ fields should be greater than those of RH neutrinos, $ M_i\gtrsim 2m_{N_{R,\alpha}}$. Taking into account the resonant effect in the CP violation parameter, we will have nearly degenerate masses of $ \varphi_{i}$s at the TeV scale. Then, using the out-of-equilibrium condition for $ \varphi_{i}$ decays, we will find $ \mathcal{L}_1$ according to Eq.\ (\ref{lep}).  

\begin{figure}[H]
	\centering
	\includegraphics[scale=0.55]{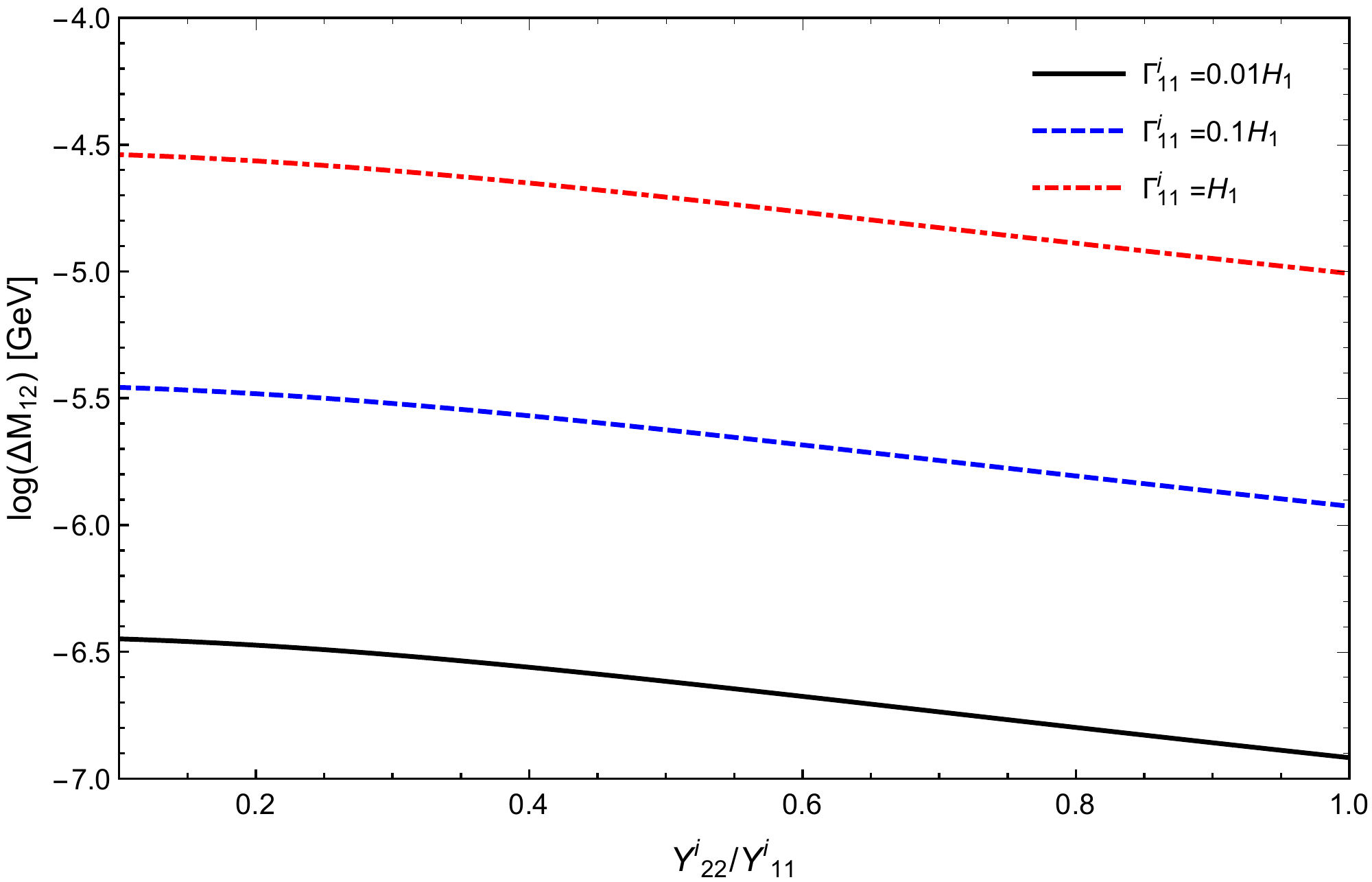}
	\caption{The required mass difference of $ \varphi_{i}$ fields versus $Y^i_{22}/Y^i_{11} $ is displayed for three different values of $ \Gamma ^i_{11}/H_1$.}
	\label{f1}
\end{figure}

\begin{figure}[H]
	\centering
	\includegraphics[scale=0.55]{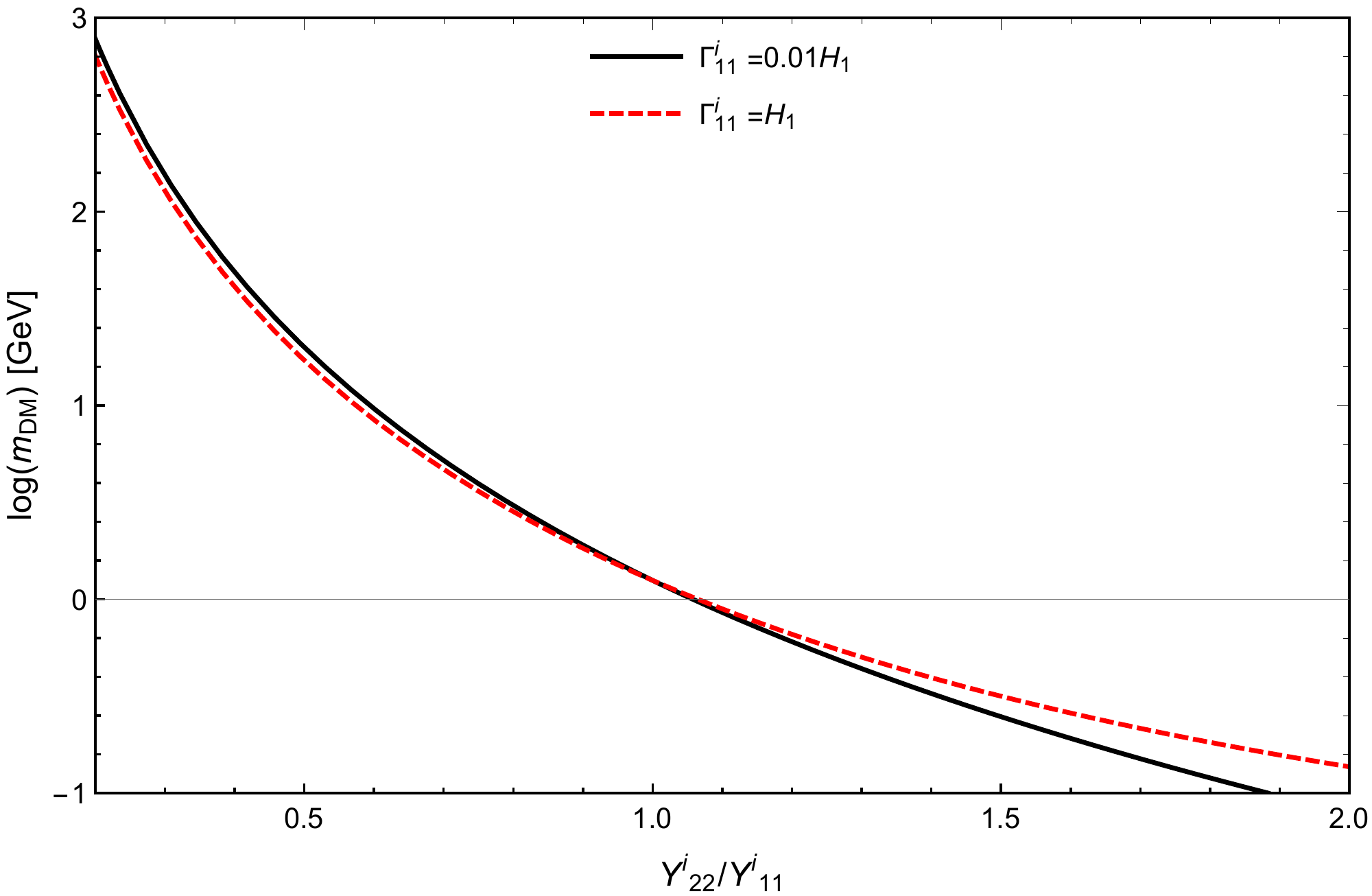}
	\caption{By changing $Y^i_{22} $, a range of DM mass is shown for two different values of $ \Gamma ^i_{11}/H_1$.}
	\label{f2}
\end{figure}

Assuming $ M_i=5\,\mathrm{TeV}$ and $ z_i=5$, we obtain $H_1\simeq 3.5\times 10^{-11}\,\mathrm{GeV} $. Moreover, assuming $\varphi_1$ and $\varphi_2$ decay similarly, we can  obtain $ \mathcal{L}_1=3\times 10^{-10}$ for $ \Gamma ^i_{11}\lesssim H_1$. Here, we take $ g_*\sim 107$ and set Yukawa couplings $ Y^i_{11}$ by three representative values of $ k_1$ where $ k_1=\Gamma ^i_{11}/H_1$. Then, through Eq.\ (\ref{lep}), we can determine $\varepsilon ^i_{11} $, as listed in Table \ref{t1}. 
\begin{table}
	\begin{center}
		\begin{tabular}{|c| c| c|} 
			\hline
			$k_1$ & $\varepsilon ^i_{11}$ & $Y^i_{11}$   \\
			\hline\hline
			1 & $5.97\times 10^{-7}$ & $5.95\times 10^{-7}$  \\ 
			\hline
			0.1 & $4.94\times 10^{-7}$ & $1.88\times 10^{-7}$ \\
			\hline
			0.01 & $4.85\times 10^{-7}$ & $5.95\times 10^{-8}$ \\
			\hline
		\end{tabular}
	\caption{The values of the CP violation parameter, $\varepsilon ^i_{11}$, and the Yukawa coupling, $Y^i_{11}$, associated with $ N_{R, 1}$ interaction are listed for three values of $ k_1=\Gamma ^i_{11}/H_1$.}
	\label{t1}
	\end{center}
\end{table}
In case the imaginary and real parts of the couplings are of the same order, Eq.\ (\ref{cp}), required $\varepsilon ^i_{11}$ values can be fulfilled for a range of values of $Y^i_{22}/Y^i_{11} $ and $ \Delta M_{12}=M_1-M_2$ which can be about $ (10^{-4}-10^{-7})\,\mathrm{GeV}$ as can be seen in Fig.\ (\ref{f1})\footnote{Another range of solutions for $ \Delta M_{12}$ can be around 11 orders of magnitude smaller, not shown in Fig.\ (\ref{f1}), e.g. for $ Y^i_{22}=Y^i_{11}$}.
On the other hand, from the following relation for the asymmetric DM, one can predict the DM mass  
\begin{equation}\label{dm}
\frac{\Omega_{\mathrm{DM}}}{\Omega_{\mathrm{B}}}=\frac{n_{\mathrm{DM}}\,m_{\mathrm{DM}}}{n_{\mathrm{B}}\, m_{\mathrm{B}}}\sim \frac{m_{\mathrm{DM}}}{m_p}\frac{\varepsilon_{i2}}{\varepsilon_{i1}}\frac{\eta_{i2}}{\eta_{i1}}\sim 5 
\end{equation}
where $m_p\sim 1\,\mathrm{GeV}$ is the proton mass and $ m_{\mathrm{DM}}\equiv m_{N_{R,2}}$. We assume $m_{\mathrm{DM}}\lesssim m_{N_{R,1}} $. Using the obtained results and Eq.\ (\ref{dm}), we can calculate $\mathcal{L}_2$ as a function of $Y^i_{22}/Y^i_{11} $. Indeed, relying on $ \Gamma ^i_{22}=\Gamma ^i_{11}Y^{\dagger}_{22}Y_{22}/(Y^{\dagger}_{11}Y_{11})$ for two values of $k_1$, we find $\mathcal{L}_2$ and thereby DM mass as a function $Y^i_{22}/Y^i_{11} $, displayed in Fig.\ (\ref{f2}). Eventually, depending on $ Y^i_{22}$ values, for example for $ k_1=0.01$ and $ 0.19\lesssim Y^i_{22}/Y^i_{11}\lesssim 20 $, a wide range of DM masses from 10 keV to TeV can be found in the scenario.

Therefore, we can find some parameter spaces of the model compatible with the observed values of the baryon asymmetry and DM.
\vspace{0.25cm}
\subsection{Low-energy DM interactions}

In this section, we discuss relevant effective interactions of the DM with SM particles at low-energy scale and estimate the cross section.

\begin{figure}
	\centering
	\subfloat{
		\centering
		\includegraphics{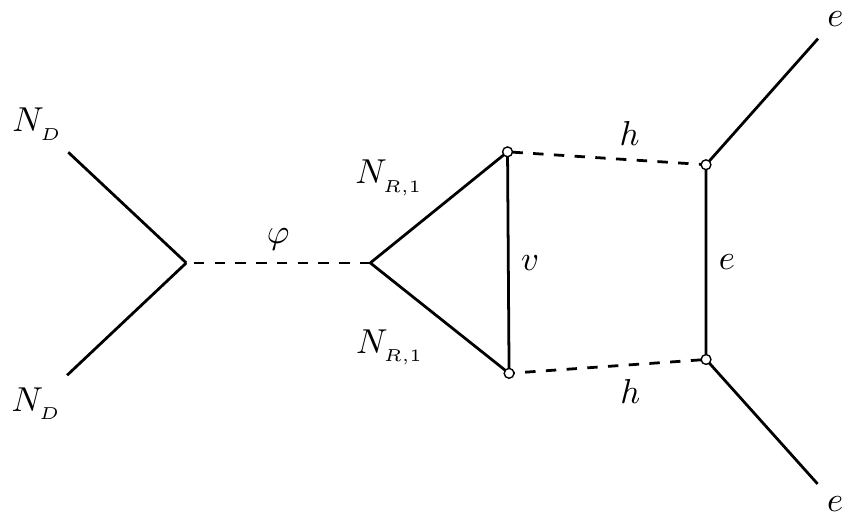} 
		\hspace{3cm}
		\includegraphics[scale=1.2]{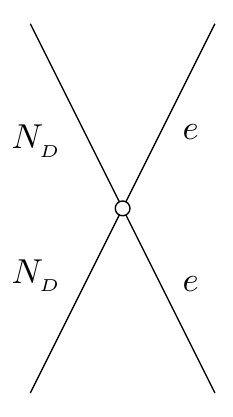}
	}
	\caption{Left: The Feynman diagram of DM-electron interaction. Right: The induced interaction at low-energy with the effective coupling. }
	\label{f3}
\end{figure}

We first study the possible DM and electron interaction which may contribute to the electron recoil events in direct detection experiments \cite{Essig:2012yx,Emken:2019tni,Shakeri:2020wvk}. The ($N_D\,e\rightarrow N_D\,e $) interaction, where $ N_D\equiv N_{R, 2}$, can be described in the loop level as represented in Fig.\ (\ref{f3}).
At low-energy regime in the lab frame, from the energy momentum conservation $ E_{N_{D_i}}+m_e=E_{N_{D_f}}+E_{e_f} $ and $ \vec{p}_{N_D}=\vec{k}_e+\vec{k}_{N_D} $. In this case and for the non-relativistic DM, we calculate approximately the total cross section as
\begin{equation}
\sigma(N_D\,e\rightarrow N_D\,e)\sim\frac{|\mathcal{M}|^2}{\pi\, m_e m_{\mathrm{DM}}}\sim\frac{\mathcal{G}_1^2}{f^4} m_{\mathrm{DM}}m_e\sim10^{-55}\,\mathrm{cm}^2\left(\frac{\mathcal{G}_1}{10^{-7}}\right)^2\left(\frac{m_{\mathrm{DM}}}{1\,\mathrm{GeV}}\right)
\end{equation}
where $\mathcal{G}_1 $ is the dimensionless effective coupling, $ f\sim 1\,\mathrm{TeV}$ and $ m_e\simeq 0.5\,\mathrm{MeV}$. For this elastic scattering, the electron recoil energy \cite{Essig:2011nj} would be $ E_r=E_{e_f}-m_e\sim (m_{\mathrm{DM}\,}v_D )^2/m_e \sim \mathrm{MeV}$ where $ v_D$ is the DM velocity and is considered around $ 10^{-3}$ due to the escape velocity from the Milky Way \cite{Smith:2006ym}. Also, different values of the parameter space $(\mathcal{G}_1, m_{\mathrm{DM}})$ can be probed by direct detection experiments such as XENONnT \cite{Aprile:2015uzo}, LZ \cite{Akerib:2015cja}, PandaX-II \cite{Cui:2017nnn} and DARWIN \cite{Aalbers:2016jon}.

\begin{figure}
	\centering
	\subfloat{
		\centering
		\includegraphics{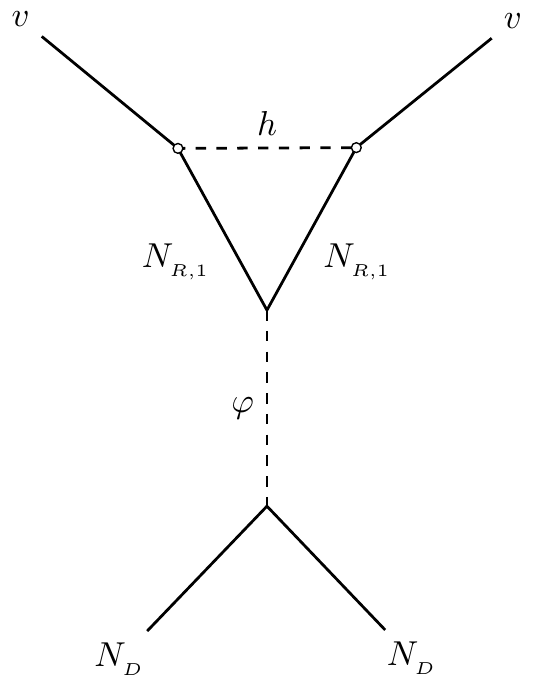} 
		\hspace{3cm}
		\includegraphics[scale=1.2]{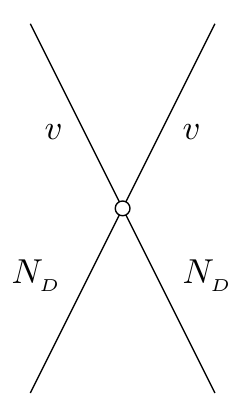}
	}
	\caption{Left: The Feynman diagram of DM-SM neutrino interaction. Right: The induced interaction at low-energy with the effective coupling. }
	\label{f4}
\end{figure}

Another interesting effective process would be the interaction between the DM and SM neutrino $ (N_D\, N_D\rightarrow v\,v)$, shown in Fig.\ (\ref{f4}).
In this case, we obtain the cross section in the center of mass frame. At low energies, considering the non-relativistic DM and $ m_v\simeq 0$, we have $E_{N_D}\sim  m_{\mathrm{DM}}$ and $ E_v\simeq |\vec{p}_{v}|\sim E_{N_D}$. Therefore, the total cross section would be
\begin{equation}
\sigma(N_D N_D\rightarrow v v)\sim\frac{|\vec{p}_{v}||\mathcal{M}|^2}{64\pi E_{N_D}^2 |\vec{p}_{N_D}|}\sim\frac{\mathcal{G}_2^2\,m_{\mathrm{DM}}^2}{8\pi f^4\,v_D}\sim 10^{-50}\,\mathrm{cm}^2 \left(\frac{\mathcal{G}_2}{10^{-7}}\right)^2\left(\frac{m_{\mathrm{DM}}}{1\,\mathrm{GeV}}\right)^2\left(\frac{10^{-3}}{v_D}\right)
\end{equation}
where $|\vec{p}_{N_D}|\simeq m_{\mathrm{DM}\,}v_D $. The parameter space including $ (\mathcal{G}_2, m_{\mathrm{DM}})$ can be investigated via DM-neutrino interaction experiments \cite{Arguelles:2019ouk}.

Moreover, based on the model, one can consider invisible decay processes of the Higgs \cite{Aaboud:2018sfi} $(h\rightarrow N_D\, N_D\, \mathrm{f}\,\mathrm{f}) $, where $\mathrm{f}$ is a SM fermion. In fact, with a Feynman diagram similar to Fig.\ (\ref{f4}), it can be shown that the cross section of such effective Higgs-DM portal $(h\,h\rightarrow N_D\, N_D) $ would not be large, $ \sigma\lesssim 10^{-47}\,\mathrm{cm}^2$, for the low DM mass, $m_{\mathrm{DM}}<m_h/2 $, and the small effective coupling  $\mathcal{G}_3\lesssim 10^{-7}$. We leave detailed calculations associated with phenomenological features of the model for a future work.

\section{Conclusion}
In this paper, we have tried to address two important issues, the baryon asymmetry of the Universe and DM problem, which cannot be explained by the SM. Using CHMs which are also interesting models addressing the hierarchy problem, we have proposed a model to explain these problems simultaneously. More precisely speaking, in a minimal CHM framework whose SM sector is extended by singlet heavy neutrinos, we have explored the possibility of a matter and DM asymmetric scenario. Indeed, such a model is motivated by the observed closeness of the dark and baryonic energy density. 

In this scenario, from the new strongly coupled system, we have considered resonant bound states which decay to RH neutrinos. In addition, one of the RH neutrinos can play the role of DM, stabilized by a $ \mathbb{Z}_2$ discrete symmetry. We obtained the decay rate of these lepton-number violating processes which should satisfy the out-of-equilibrium condition. Furthermore, as an interesting feature, the required CP violation can be sufficiently provided in the model due to the resonant effect of bound states. We then have calculated the number densities by solving the Boltzmann equations. The generated asymmetry in the RH neutrino sector is induced to the SM leptons by their interaction with visible RH neutrinos and eventually via sphalerons it is converted to the baryon asymmetry. 

By a numerical analysis, we have shown the observed baryon asymmetry and relic abundance of DM can be achieved at the compositeness scale for TeV scale bound states. As another merit which can be also experimentally interesting, depending on the strength coupling of the DM interaction with bound states, a range from 10 keV to TeV for the DM mass can be found in the model. Eventually, we discussed possible low-energy DM interactions with SM particles and associated experiments which may probe the parameter space of the model.

\vspace{2cm}
\textbf{Acknowledgment}\\\\
I would like to thank Soroush Shakeri and Majid Ekhterachian for helpful comments and discussions.

%\newpage
	
%%%%%%%%%%%%%%%%%%%%%%%%%%%%%%%%%%
\end{document}